# Title Page

# Nitro-compounds and GHG exhaust emissions of a pilot diesel-ignited ammonia dual-fuel engine under various operating conditions

**Run Chen[1,2], Tie Li[1,2*], Xinran Wang[1], Shuai Huang[1], Xinyi Zhou[1], Shiyan Li[1], Ping Yi[1,2]**

[1] State Key Laboratory of Ocean Engineering, Shanghai Jiao Tong University

[2] Institute of Power Plants and Automation, Shanghai Jiao Tong University

*Corresponding Address: 800 Dongchuan Rd., Shanghai, PR China, 200240, Tel.: (86)21-3420-8348; E-mail: litie@sjtu.edu.cn

**SYNOPSIS**

Ammonia is the most promising green fuel, but few reports on emissions characteristics of ammonia-fueled engines. This study clarifies the effects of ammonia used as a power source on the atmospheric contamination.

**ABSTRACT:** In the transportation sector, ammonia used as a power source plays a significant role in the scenario of carbon neutralization. However, the engine-out emissions correlations of ammonia-diesel dual-fuel (DF) engines are still unclear, especially the nitro-compounds of great concern and GHG. In this study, the engine-out emissions are evaluated by using a four-cylinder ammonia/diesel DF engine. Various operating conditions consisting of ammonia energy ratio (AER), engine load, and speed were carried out. Unburned $NH_3$ increases with raising ammonia content but decreases with increasing engine load and speed. The $NO+NO_2$ tendency shows a non-linearity trend with increasing ammonia content, while a trade-off correlation is linked to $N_2O$. The $N_2O$ emission of ammonia engine significantly weakens the beneficial effect of GHG reduction, the 30% and 50% decarbonization targets need at least 40% and 60% ammonia energy without regard to $N_2O$'s effect, while at least 65% and 80% ammonia energy respectively with considering $N_2O$. $N_2O$ presents a parabolic-like tendency with AERs. Advanced pilot-diesel injection timing helps to reduce both $NH_3$ and $N_2O$, but this effect becomes insignificant as the AER is less than 0.4. A combustion strategy of the rapid heat release and ammonia-governed heat release respectively are revealed.

## INTRODUCTION

The 99% of transportation power source is dominated by high-carbon fossil fuels, such as diesel, gasoline, and heavy oil fuel, which are greatly far off the decarbonization requirements, especially for heavy-duty vehicles and ocean-going shipping. The International Maritime Organization (IMO) proposed a target of 30% and 50% lowing carbon emissions by 2030 and 2050 respectively [1]. Such an urgent decarbonization mission is widely recognized to be perfectly resolved by using alternative fuels with low/zero carbon.

Many literatures and reports [2-7] have reviewed the potential green fuels. Ammonia is easily to be liquefication under -33°C @1atm or 0.9MPa@room temperature, which is more convenient compared with LNG and hydrogen. In addition, ammonia with 17.5wt% of hydrogen helps to contribute the clean combustion as a high-efficiency hydrogen carrier. However, the ammonia's weaknesses of higher ignition resistance and lower flame propagation speed are prone to narrow the reliable operating window, leading to either misfire or undesirable engine-out emissions. The worsened exhaust performance is expected to greatly impede ammonia as a power source in internal combustion engines (ICEs) [8-9].

Due to the convenient utilization and accessibility of diesel fuel, many scholars have focused on the pilot diesel-ignited ammonia dual-fuel CI (compression ignition) engines for recent 15 years. Reiter et al. [10-11] used pre-mixed ammonia in a 4-cylinder CI engine and found the indicated thermal efficiency (ITE) decreased proportionally by adding ammonia. Niki et al. [12-14] investigated the low-pressure gaseous ammonia injection utilized in a CI engine. They pointed out that the early diesel injection helped to reduce both unburned $NH_3$ and $N_2O$. Recently, Yousef et al. [15-16] found that the thermal $DeNO_x$ process helped to decrease $NO_x$ if the ammonia energy share was less than 40%. Zhou and Li et al. [17] compared the low-pressure premixed and high-pressure spray combustion modes of a low-speed marine engine by numerical simulation. Jin et al. [18] optimized the diesel pilot injection strategies in the ammonia dual-fuel engine by numerical simulation and achieved a 14% reduction of $CO_2$. As mentioned above, lots of studies have laid concentration on the energy ratio of ammonia and numerical simulation. However, the engine-out emissions of unburned $NH_3$ and nitrogen oxides (namely nitro-compounds), and $CO_2$ under different engine operating conditions are still unclear, especially since the correlations between the emissions are absent. Therefore, this study was experimentally conducted by a 4-cylinder ammonia-diesel dual-fuel engine to give an insight into the engine-out emissions under various operating conditions. Finally, the possible improvements in engine-out emissions are discussed.

**EXPERIMENTAL SETUP AND PROCEDURES**

**Experimental Setup**

In this study, the 4-cylinder and turbocharging CI engine modified by adding gaseous ammonia injection was employed. The specification of the engine system was listed in Table 1. An 8-hole fuel injector for each cylinder provided a pressurized diesel to approach the pilot-ignited premixed ammonia combustion. The diesel quantity was measured by the fuel meter (AVL List GmbH, PD735s). The single-point injection of ammonia was used at the intake system in this study. The liquified ammonia was supplied by the cylinder and gasified through the heat exchanger at the downstream ammonia fuel line. The gaseous ammonia was controlled by over 30°C and fine-regulated to about 0.6MPa (abs.) which ensures the stable gaseous ammonia state. A surge tank was used to eliminate the supply pressure oscillation and the mass flow controller (Brooks Instrument LLC, SLA5800 Series) was employed at the terminal of the ammonia fuel line to precisely control the target energy ratio. The intake temperature was set by the intercooler system with an accuracy of ±2°C.

**Table 1.** Specification of the engine

| | |
|---|---|
| Engine Type | 4-cylinder, Turbocharged |
| Engine configuration | Four-stroke, direction injection |
| Valves per Cylinder | 4 |
| Bore × Stroke [mm] | 95 × 102 |
| Compression ratio [-] | 17.5 |
| Nozzle hole diameter [mm] | 0.127 |
| Nozzle number | 8 |
| Power @full load [kW] | 22.4@1000 rpm |
| Rated Power [kW] | 112@2800 rpm |
| Displacement [L] | 2.89 |

The in-cylinder pressure was measured by a highly precise piezoelectric pressure sensor (AVL List GmbH, GH15DK) with a crank angle resolution of 0.5°CA. The sensor's thermal sensitivity is less than ±2% in the temperature range of 20°C to 400°C. The signals of intake/exhaust static pressures and temperatures were collected through the engine central controlling system built by Labview. The exhaust gas analyzer (Horiba Ltd., Mexa-one Plus) was applied to sampling and quantitively measured the conventional engine-out emissions consisting of the total unburned hydrocarbon (THC), $NO_x$, $CO_2$, and so on. And the special pollutants, such as the unburned $NH_3$ and $N_2O$, etc. were collected by the Fourier-transform infrared gas analyzer (Horiba Ltd., FTX-ONE-CS) with an accuracy of ±1.0%.

**Experimental Conditions and Methodology**

Table 2 shows the engine operating conditions. The ammonia injected quantities were set to the energy shares in the range of 0% to 90% (namely 0%e to 90%e), aiming to investigate the effects of ammonia on the dual-fuel mode at the 75% load with the diesel injection timing of the maximum break torque (MBT). Also, the sweeping of the diesel injection timing was conducted to investigate the emission characteristics. Under the 80%e condition, the effects of engine load from 50% to 100% on the engine-out emissions were clarified. Due to the possible ammonia misfire caused by the over-lean at high speed, the ammonia energy ratio was set to 60%e to explore the impacts of engine speed on the pollutant emissions of ammonia-diesel dual-fuel mode.

**Table 2.** Experimental conditions

| Experimental Tests | Parameters | | | |
|---|---|---|---|---|
| | Diesel Inj. Timing | Ammonia Energy Ratio | Engine Load | Engine Speed |
| Engine speed [rpm] | 1000 | 1000 | 1000 | 800 to 2500 |
| BMEP [MPa] | 0.7 | 0.7 | 0.5, 0.7, 0.8, 0.9 | 0.7 |

| Diesel Injection Pressure [MPa] | 120 | 120 | 120 | 120 |
| --- | --- | --- | --- | --- |
| Ammonia energy share [%e] | 0, 20, 40, 60, 80, 90 | 0, 20, 40, 60, 80, 90 | 80 | 60 |
| Diesel Injection Timing [°CA aTDC*] | Sweep MBT±4°CA if w/o misfire or CoV>5% | MBT | MBT | MBT |

*aTDC: after top dead center

In this study, the apparent heat release rate (AHRR) was derived from the in-cylinder pressure measurement as shown in Eq.(1), where $\gamma$ is the specific heat ratio calculated based on the two-zone mode. $P$ is the cylinder pressure, $V$ is the cylinder volume, $\theta$ is the crank angle.

$$AHRR = \frac{\gamma}{\gamma-1} \cdot P \cdot \frac{dV}{d\theta} + \frac{1}{\gamma-1} \cdot V \cdot \frac{dP}{d\theta} \qquad (1)$$

The ammonia energy ratio (AER, e%) was determined by the total heating value of injected ammonia divided by the total heating value of the fuel amount as shown in Eq.(2).

$$AER(e\%) = \frac{Q_{NH3}}{Q_f} = \frac{m_{NH3} \cdot LHV_{NH3}}{m_d LHV_d + m_{NH3} \cdot LHV_{NH3}} \qquad (2)$$

where $Q_{NH3}$ is the injected ammonia energy. $Q_f$ is the total fuel energy. $m_{NH3}$ and $m_d$ is the mass flow of ammonia and diesel respectively. $LHV_{NH3}$ and $LHV_d$ is the lower heating value of ammonia and diesel. The combustion efficiency ($\eta_c$) was calculated by the heat release by the combustion ($Q_r$)) and the total energy of input fuels, as expressed in Eq.(3).

$$\eta_c = \frac{Q_r}{Q_f} \times 100 \qquad (3)$$

The stable combustion limits were defined as the either misfire or the correlation of variation (CoV) for indicated mean effective pressure (IMEP) over 5%.

**RESUTLS AND DISCUSSION**

**Unburned Ammonia Slip**

Figure 1(a) shows the unburned $NH_3$ slip and unburned loss varied with the AER. The unburned loss was calculated based on the energy of the engine-out emissions, including CO, $CO_2$, THC, and unburned $NH_3$. There is a share of approximately 8.5% at the unburned loss under the 90%e condition. This is attributed to the higher ammonia content causing the lower flame propagation speed. The laminar burning velocity of ammonia at near stoichiometric condition is only about 10% of methane [19]. The low flame speed leads to an increased amount of unburned $NH_3$. It should be noticed that the share of unburned $NH_3$ mass slip to the total $NH_3$ fuel mass is almost constant (about 10%) for all AERs, as shown in Fig.1(b). This feature is very useful to predict the unburned $NH_3$ emission as the ammonia content alters but is greatly dependent on the engine chamber design. Reiter et al. [11] pointed out that the ammonia combustion efficiency sustains a constant (about 5%) under different AERs and Niki et al. [20] also reported a nearly 15% of the unburned $NH_3$ as increasing the $NH_3$ flow rate from 0 to 13.3L/min. Theoretically, the unburned $NH_3$ is considered to produce from the piston crevice and incomplete combustion. In this study, the increase of in-cylinder pressure for the operating conditions (from 0%e to 80%e) at MBT has few impacts on the unburned $NH_3$ emissions as shown in Fig.1(b), indicating that the unburned $NH_3$ in the piston crevice seems to unaffected by the cylinder pressure. This result is consistent with that of Ref.[20]. As a consequence, the incomplete combustion including flame quenching and inaccessible regions of the diesel flame mainly contributes to the unburned $NH_3$. According to the discussion above, the effects of heat transfer characteristics of $NH_3$ premixed flame and spatial penetration of diesel flame plumes on the ammonia-diesel duel-fuel mode should be carefully examined further.

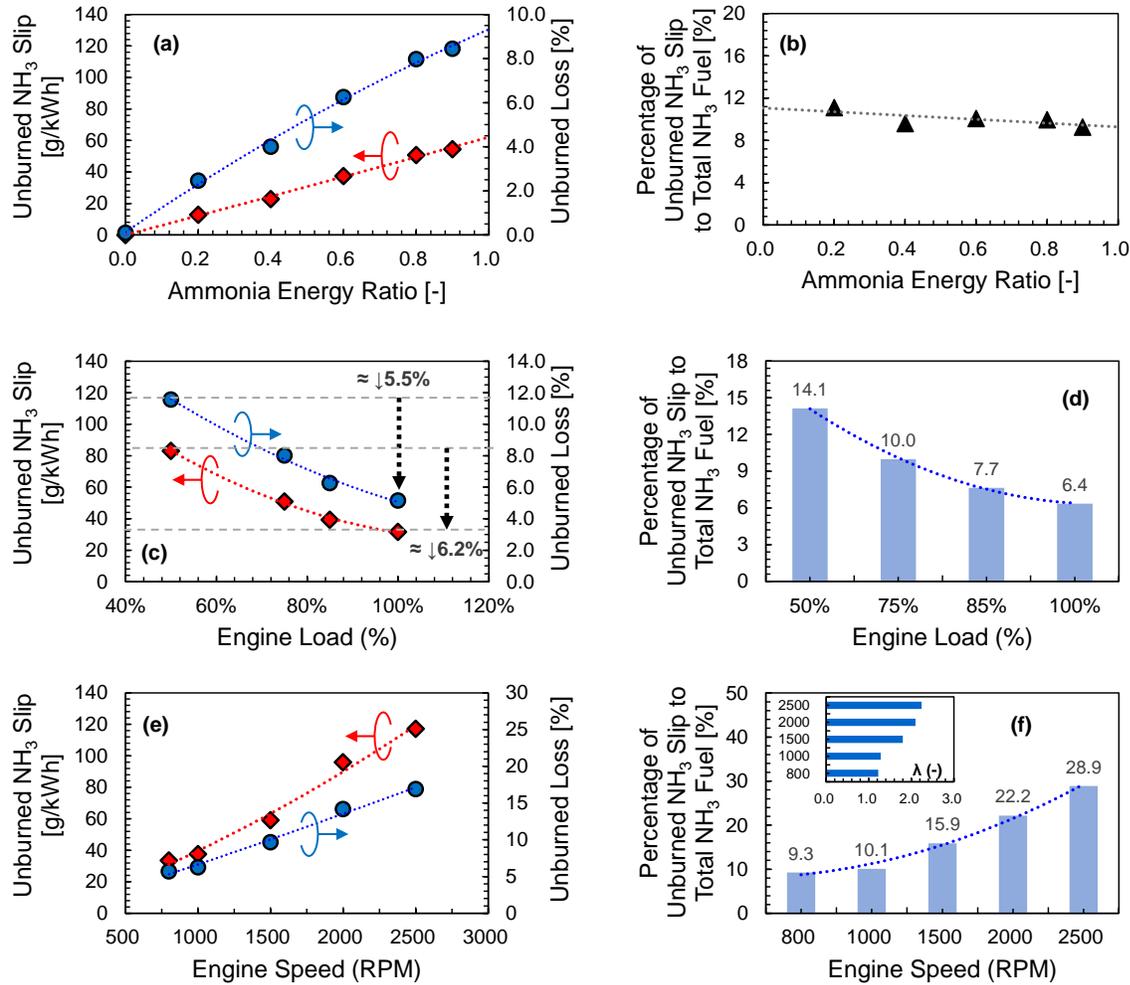

**Figure 1.** Unburned NH$_3$ slip characteristics under various operating conditions. (a-b) Unburned NH$_3$ vs. AER @75% engine load/1000rpm/MBT; (c-d) Unburned NH$_3$ vs. engine load @80%e/1000rpm/MBT; (e-f) Unburned NH$_3$ vs. engine speed @80%e/75% engine load/MBT

Figure 1(c) shows the tendencies of unburned NH$_3$ and unburned loss under different engine loads. The unburned NH$_3$ decreases by about 6.2% as the engine load increases from 50% to 100%. The higher in-cylinder pressure and temperature under high load promote ammonia decomposition and improve the combustion speed, thus decreasing the unburned NH$_3$. This facilitation also suggests that 5.5% decrease in the unburned loss from low to high load. Such a tendency indicates that for the ammonia-diesel dual-fuel mode at high AERs, the unburned NH$_3$ slip nearly determines the total unburned loss.

Under the high engine load, the share of unburned $NH_3$ to the total $NH_3$ fuel amount decreases significantly as presented in Fig.1(d). The unburned $NH_3$ under the low-/mid loads should be taken carefully since these loads are the common operating conditions for ocean-going vessels. Especially the 50% load leads to an over twice increase of the unburned $NH_3$ share compared with that of 100% load, which might become a great potential challenge for ammonia-power vessels to fitting the future coastal emission control regulations. As shown in Fig.1(e). The unburned $NH_3$ emission increases about 2.5 times from 800rpm to 2500rpm, as well as the unburned loss rises nearly 12%. The possible reason is attributed to the mixture being lean with increasing engine speed, as shown in Fig.1(f). The mixture becomes over lean (lambda = 2.2 at 2500rpm) at the high speed due to the increase of the air amount. Over lean mixture has a great impact on the ammonia flame quenching during the combustion.

**$NO_x$ ($NO+NO_2$) Emissions**

In Fig.2(a), the NOx (represented by $NO+NO_2$) tendency shows a non-linearity varied with AERs at MBT. When the AER is less than 60%e, the $NO_x$ of dual-fuel mode is lower than diesel-only mode, suggesting that the reduction $NO_x$ can be achieved under a lower ammonia energy ratio. From 20%e to 60%e, the $NO_x$ ($NO+NO_2$) decreases about 20% compared with that of pure-diesel mode (donated by the blue region). The reduction of $NO+NO_2$ under lower AER is attributed to the suppression of thermal NO production by lowing the diesel content and reduced combustion temperature. Nevertheless, when the AER reaches over 60%, the $NO+NO_2$ emission amount increases dramatically. At 80%e and 90%e conditions, the $NO+NO_2$ emissions show 46% higher than that of the pure-diesel mode. More ammonia content facilitates the increase of HNO from $NH_3$ pyrolysis, accelerating the important fuel NO route as $HNO + O_2 + M = NO$. As a result, the fuel-borne NO trends to become the dominance of $NO_x$ production as the ammonia content approaches a very high level.

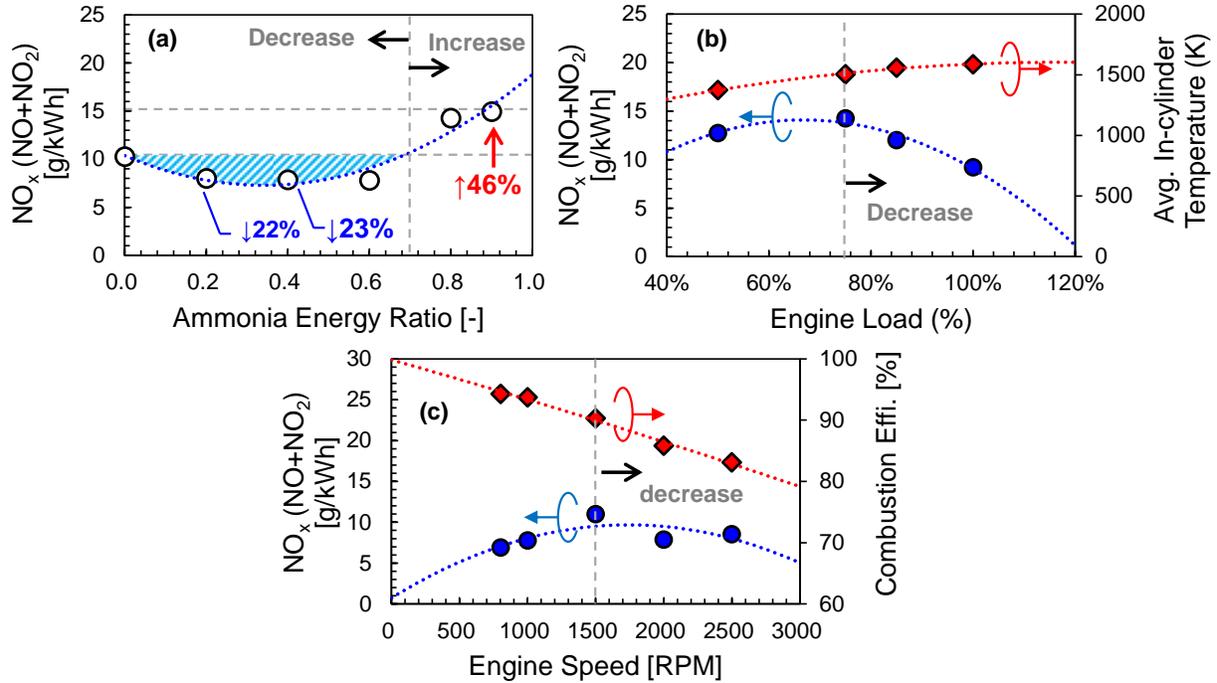

**Figure 2.** $NO_x$ (NO+$NO_2$) variation with different operation conditions at MBT. (a) $NO_x$ (NO+$NO_2$) vs. AER @75% engine load/1000rpm/MBT; (b) $NO_x$ (NO+$NO_2$) vs. engine load @80%e/1000rpm/MBT; (c) $NO_x$ (NO+$NO_2$) vs. engine speed @80%e/75% engine load/MBT

Figure 2(b) shows the $NO_x$ (NO+$NO_2$) tendency varied with different engine loads under 80%e and 1000rpm. The average in-cylinder temperatures derived from SOI to CA90 also are presented. The $NO_x$ emission firstly increases as the engine load rises, but decreases when the load exceeds 80%. As the engine load raises from 40% to 75%, the $NO_x$ increases probably as a result of the thermal $NO_x$ significantly produced due to the in-cylinder temperature rising. Even though the DeNO$_x$ chemistry through the pathways of $NH_2 + NO = N_2 + H_2O$ and $NH_2 + NO = NNH + OH$ is activated from 1100 to 1400K [21-22], the total NO+$NO_2$ emissions from 40% to 75% load is determined by the balance of local thermal $NO_x$ formation and DeNO$_x$ reactions during combustion. Under the high load condition, however, the $NO_x$ turns to decrease obviously although the in-cylinder temperature increases steadily. This is possibly attributed to more $NH_3$ burning under high loads, contributing to acceleration of the thermal $DeNO_x$. The effects of engine speed on the $NO_x$ emission are exhibited in Fig.2(c). The reduction of combustion efficiency with increasing engine speed suggests that the burning intensity is suppressed by raising speed. In addition, the leaner in-cylinder mixture also promotes reducing the combustion temperature, so that the thermal NO formation is impeded with increasing engine speed.

**$N_2O$ Emission**

The feature of $N_2O$'s greenhouse-warming potential (GWP) of 300 times than $CO_2$ greatly determines the net effect of shipping decarbonization by using ammonia as the power source. Figure 3(a) shows the $N_2O$ variation with AERs, combined with the average in-cylinder temperature. The $N_2O$ increases with the ammonia energy taking up to 60% and then reduces when the AER becomes higher in further. First, the $NH_2$ radicals from the $NH_3$ pyrolysis fast dehydrogenated and facilitate the $N_2O$ formation by $NH + NO = N_2O + H$ as the temperature is over 1400K [23-24], which is considered as the primary path for $N_2O$ formation [25]. Secondly, Mathieu et al. [26] also pointed out that the $N_2O$ thermal consumption by $N_2O + H = N_2 + OH$ started as the ambient temperature was above 1650K. The above reasons lead to the increase of $N_2O$ emissions from 20%e to 60%e. When the AER reaches over 0.8, the $N_2O$ significantly decreases due to the main consumption of $N_2O$ through $N_2O + M = N_2 + O + M$ presenting under lower in-cylinder pressure while the temperature is over 1300K [27]. Although the in-cylinder temperature reduces significantly (below 1400K) at AER=0.9 as highlighted by a red box in Fig.3(a), the $N_2O$ emission maintains a similar level to AER=0.8. The $N_2O$ is reported to be produced through $NH_2 + NO_2 = N_2O + H_2O$ at the lower temperature range of 970 to 1300K [28]. Combined with the above discussion, it indicates that a stagnation of $N_2O$ formation might present between 1300 to 1400K, possibly leading to a similar level of N$_2$O emissions for AER=0.8 and 0.9.

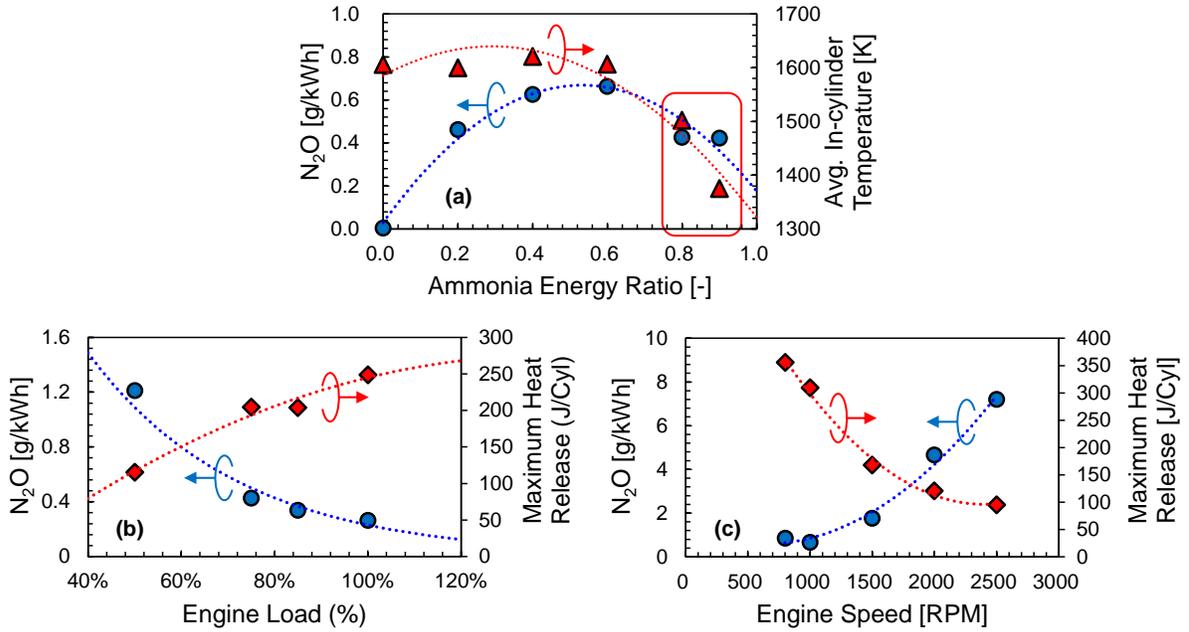

**Figure 3.** N$_2$O variation with different operation conditions at MBT. (a) N$_2$O vs. AER @75% engine load/1000rpm/MBT; (b) N$_2$O vs. engine load @80%e/1000rpm/MBT; (c) N$_2$O vs. engine speed @80%e/75% engine load/MBT.

Figures 3(b) and (c) exhibit the N$_2$O variation with different engine loads and speeds for AER=0.8 at MBT respectively. Since the in-cylinder temperature and pressure increase at higher engine load conditions, the maximum heat release raises, significantly promoting the N$_2$O thermal decomposition thus reducing N$_2$O emissions as shown in Fig.3(b). Moreover, the local leaner mixture as the engine speed increases, leads to a slower flame propagation speed. This impact results in a lower heat release (as shown in Fig.3(c)) but also facilitates N$_2$O formation under low-temperature conditions [28-29].

## CO$_2$ and CO$_{2e}$ Emissions

Figure 4(a) shows the $CO_2$ reduction varied with the AER. The equivalent $CO_2$ emission, namely $CO_{2e}$, is calculated by the $CO_2$ emission adding the $N_2O$ amount multiplied by 300 times. The $CO_2$ reduction comes up to 31% when the AER reaches 0.4, nearly 50% as AER=0.6, without considering the effects of $N_2O$. Theoretically, the 60% energy share of ammonia in the dual-fuel mode can achieve 50% decarbonization. However, the true carbon reduction is greatly determined by the $N_2O$ emission as a result of $N_2O$'s significant greenhouse effect as shown by a red zone in Fig.4(a). The ammonia energy share has to reach at least 80% to fit the 50% $CO_2$ reduction, indicating that the decarbonization requirements become more rigorous dramatically if considering the $N_2O$ emission. Due to the peak of $N_2O$ emission occurring at the AER=0.4 to 0.6, therefore, there is an observable decarbonizing weakened region (donated by red color) at the above conditions, which dramatically narrows the true decarbonization (marked by green color) operating window under different AERs. As a result, the true 50% decarbonization is approached only the ammonia energy share is up to 0.8. As a result, a disruptive technology is needed to solve the problem of high $N_2O$ emission of ammonia engine.

Figure 4(b) shows the $CO_2$ reduction under different engine loads at AER=0.8 and MBT. There is an absent difference for the direct measurement of $CO_2$ emission (about 200g/kWh) under various engine loads. As a result, the equivalent $CO_2$ emission is determined by the $N_2O$ formation. Under the 40% engine load, a large amount of $N_2O$ is produced, probably as a result of relatively low-temperature combustion. This leads to a very high $CO_2e$ emission of nearly twice those of other engine loads. In addition, the increased engine speed results in increasing $CO_2e$ emission as shown in Fig.4(c), which exhibits a similar trend to the $N_2O$ emission in Fig.4(c).

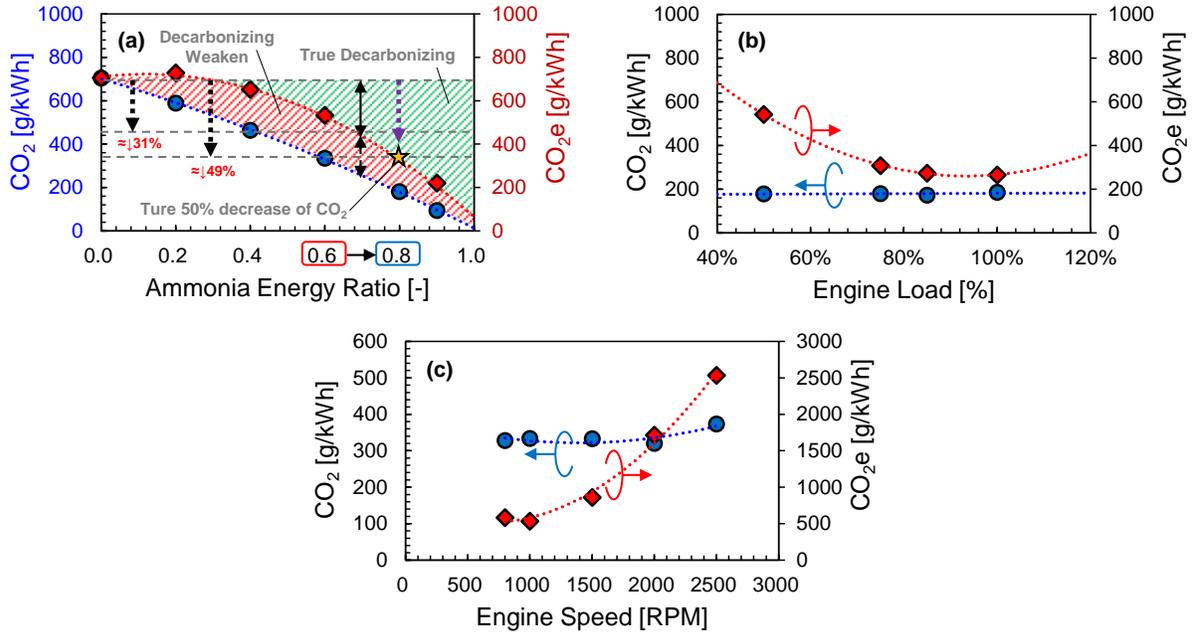

**Figure 4.** $CO_2$ variation with different operation conditions at MBT. (a) $CO_2$ vs. AER @75% engine load/1000rpm/MBT; (b) $CO_2$ vs. engine load @80%e/1000rpm/MBT; (c) $CO_2$ vs. engine speed @80%e/75% engine load/MBT.

## Combustion Decoupling Characteristics of Ammonia-diesel Duel-fuel Mode

Figure 5 shows the 80%e condition compared with the 20% diesel-only (DO) mode at the same engine speed and engine load. As reported in Ref.[10], the combustion phase shifted advanced if the diesel injection timing is kept the same for the DO mode. To make the comparison more conveniently, the diesel injection timing for the DO mode is retarded by 2 °CA to maintain nearly the same rapid pressure rising with the DF mode. The diesel injection retardation causes about a1.4% error in the indicated work, which is small enough to be neglected in this study.

In Fig.5(a), the peak of in-cylinder pressure is significantly retarded when the ammonia is introduced into the mixing fuel, due to the ammonia's lower flame speed increasing the ignition delay. And the peak of in-cylinder pressure increases by about 50% of 80%e condition than that of DO mode. The apparent heat release rate (AHRR) of 80%e DF mode can be separated into three periods, namely ignition delay (Start of diesel injection, SOI to CA01), rapid heat release (CA01 to CA50), ammonia-governed heat release (CA50 to CA90) as exhibited in Fig.5(b) to compare with that of the DO mode. After the ignition delay, the AHRR of DF mode rises quickly and reveals a peak of 30% higher than that of DO mode at nearly the same crank angle. The AHRR of the $NH_3$-only mode can be calculated by the DF AHRR removing that of the DO mode. During the rapid heat release period, the diesel at DO mode almost is burned out, while the ammonia content in DF mode provides about 56% combustion energy suggesting that 1) both the diesel diffusion flame and ammonia premixed combustion are present; and 2) the diesel and ammonia supply the similar level of the heat release during this period at the 80%e condition.

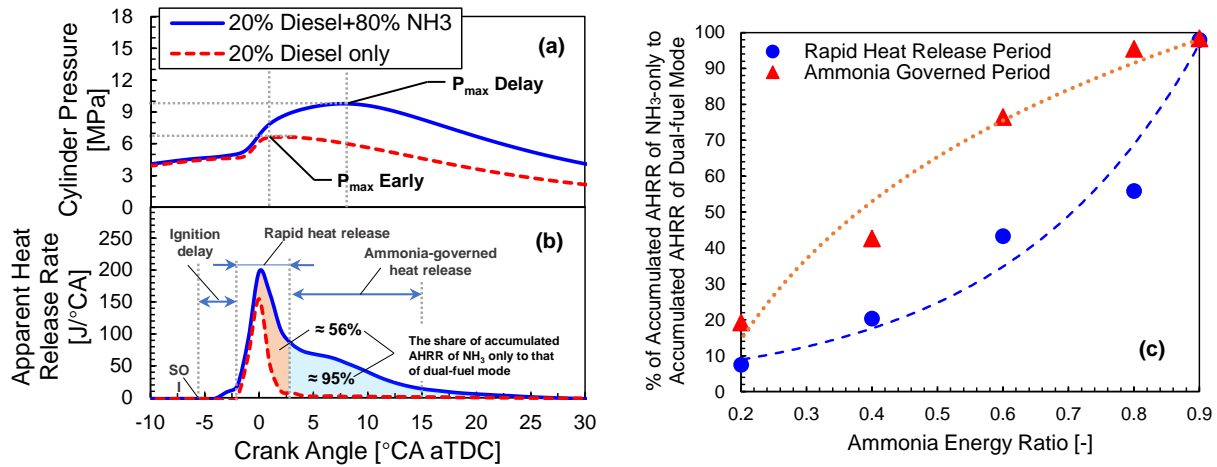

**Figure 5.** Schematic of combustion decoupling for ammonia-diesel mode. (a-b) Definition of rapid heat release period and ammonia-governed heat release period @80%e/1000rpm/75% engine load/MBT; (c) The tendency of the accumulated AHRR of $NH_3$-only mode to that of dual-fuel mode @20%e to 90%e/1000rpm/75% engine load/MBT.

After the rapid heat release period, the combustion is almost governed by the ammonia-premixed combustion in the DF mode as shown in Fig.5(b). During this period, premixed ammonia combustion is responsible for about 95% of heat release. This indicates that under higher AER conditions, the method that the proportion of ammonia heat release reduces in the ammonia-governed period while increases in the rapid heat release period may play a very significant role to improve the thermal efficiency and decrease the unburned $NH_3$.

Figure 5(c) presents the heat release contributions of $NH_3$ at different combustion periods, which are derived from the percentage of accumulated AHRR of $NH_3$-only mode to that of dual-fuel mode. A strong logarithmic relation between the heat release contribution of $NH_3$ and the ammonia share ratio is observed for the ammonia controlling period, while a significant exponential correlation is found for the rapid heat release period. Under the lower AER conditions, the larger increased rate of $NH_3$ heat contribution at the ammonia-governed period indicates that the sensitivity of ammonia heat release to the total heat release is more dramatical at the ammonia-governed period. On the opposite, the tendency of $NH_3$ heat contribution at the rapid heat release period reveals a fact that the $NH_3$ contribution plays a crucial role under the higher AER conditions during such a period. According to the discussion above, the combustion optimization at the ammonia-governed period for lower AERs, and at the rapid heat release period for higher AERs respectively are in favor of effectively improving the integrated performance of the ammonia-diesel dual-fuel mode.

If the AER raises, the increase of gaseous ammonia volume readily takes up more space in the intake port. This effect greatly decreases the charging efficiency, e.g. the volumetric efficiency reduces by about 18% from AER=0 to 0.9 in this study. The decrease in charging efficiency will deteriorate the thermal efficiency, thus worsening the engine-out emissions. The liquid ammonia with flash-boiling spray [30] directed into the intake port may play a significant role in the improvement of the volumetric efficiency, as a consequence of the large latter heat of ammonia. However, the wall impingement and interaction with the strong intake crossflow should be carefully examined if using the port injection of liquid ammonia spray.

Figures 6(a) and (b) show the correlation between the unburned $NH_3$, $N_2O$, and $NO_x$ under different AERs and pilot-diesel SOI timings. Similar levels of unburned $NH_3$ with varied diesel

SOI timing are observed from 20%e to 40%e but show an approximate logarithm relation with N$_2$O emission from 60%e to 90%e. This tendency indicates that the potential of decreasing both unburned NH$_3$ and N$_2$O simultaneously under the very high ammonia condition is possible to be achieved if the pilot-diesel SOI is advanced. But this impact becomes insignificant as the ammonia energy share reduces less than 40%. In addition, a power relation between the NO$_x$ (NO+NO$_2$) and N$_2$O emissions presents significantly as revealed in Fig.6(b), suggesting the trade-off factor exists between them. The MBT or near MBT positions show the potential to cut down the NO+NO$_2$ and N$_2$O emissions simultaneously. Moreover, under a certain range of NO+NO$_2$, the emission of N$_2$O seemingly can maintain a relatively stable level without affecting by the increasing AERs.

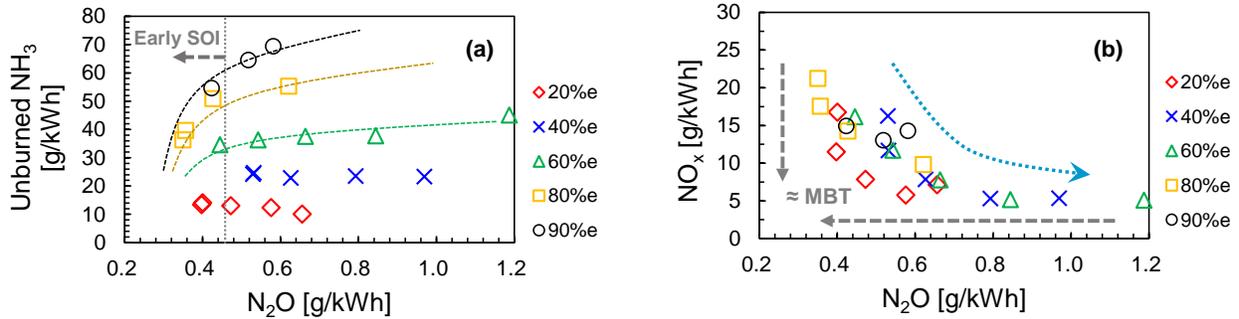

**Figure 6.** Correlation of N$_2$O, unburned NH$_3$ and NO$_x$ under different ammonia energy ratios and diesel SOI timings @1000rpm/75% engine load/MBT.

To lower unburned NH$_3$, NO$_x$ as well as N$_2$O by pilot-diesel injection, the method of firstly holding the target NO$_x$ (NO+NO$_2$), then sweeping the diesel injection timing to find the preferred SOI corresponding to the lower N$_2$O and unburned NH$_3$ might be better to acquire the compromised engine-out emissions. The multiple injection strategy integrated with the optimized pilot-diesel injection energy and dwell timing might be expected one of the most efficient ways.

One should be noticed that the N$_2$O cannot be neglected due to its very significant greenhouse effect which will greatly weaken the integrated decarbonization results, even though the N$_2$O emission level is an order of magnitude smaller than that of NO+NO$_2$ and unburned NH$_3$. N$_2$O is produced during the ammonia combustion as well as the late oxidation process of unburned NH$_3$ from engine crevices during expansion stroke [19, 31-33]. Since the SCR system also will produce a certain level of N$_2$O, it needs to take into full consideration of the N$_2$O aftertreatment position coupling with SCR, diesel oxidation catalytic (DOC), and ammonia slip catalyst (ASC). Generally, a disruptive technology is needed to solve the problem of high N$_2$O emission of ammonia engine.

**AUTHOR INFORMATION**

Corresponding Author:

*E-mail: litie@sjtu.edu.cn

**DECLARATION OF COMPETING INTEREST**

The authors declare no competing financial interest.

**ACKNOWLEDGMENTS**

The authors would like to gratefully acknowledge the support of the Major International (Regional) Joint Research Project of the National Natural Science Foundation of China (52020105009), and the National Natural Science Foundation of China (52271325) and (52171314).

**ABBREVIATIONS**

| | |
|---|---|
| AHRR | apparent heat release rate |
| AER | ammonia energetic ratio |
| aTDC | after top dead center |
| ASC | ammonia slip catalytic |
| BMEP | brake mean effective pressure |
| °CA | crank angle degree |
| CI | compression ignition |
| CA01,50,90 | crank angle of 1% / 50% / 90% mass burned fraction |
| CoV | coefficient of variation |
| $CO_{2e}$ | equivalent $CO_2$ emission |
| DOC | diesel oxidation catalytic |
| GHG | greenhouse gas |
| GWP | global warming potential |
| ICE | internal combustion engine |
| IMEP | indicated mean effective pressure |
| IMO | international Maritime Organization |
| ITE | indicated thermal efficiency |
| MBT | maximum brake torque |
| RAHR | rate of accumulated heat release |
| SCR | select catalytic reduction |
| TDC | top dead center |

| $\eta_c$ | combustion efficiency |
|---|---|
| $W_i$ | indicated work |